\documentclass[prl,twocolumn,nofootinbib,showpacs]{revtex4}
\usepackage{graphicx}% Include figure files

\usepackage{bm}% bold math
\usepackage{amssymb}
\usepackage{epstopdf}
\usepackage{amsmath}
\usepackage{amstext}
\usepackage{latexsym}
\usepackage{pstricks}

\begin{document}

\title{Structural phase transitions in low-dimensional ion
crystals} \author{Shmuel Fishman,$^1$ Gabriele De Chiara,$^{2,3}$
Tommaso Calarco,$^{4,5}$ and Giovanna Morigi$^2$} \affiliation{
$^1$ Department of Physics, Technion, 32000 Haifa, Israel\\
$^2$ Grup d'Optica, Departament de Fisica, Universitat Autonoma de
Barcelona, 08193 Bellaterra, Spain\\
$^3$ BEC-CNR-INFM \& Physics
Department, University of Trento, Via Sommarive 14, I-38050 Povo
(TN) Italy\\
$^4$ ITAMP, Harvard Smithsonian Center for Theoretical Atomic,
Molecular, and Optical Physics, Cambridge, MA, USA\\
$^5$ Abteilung Quanteninformationsverarbeitung, Universit\"at Ulm,
Albert-Einstein-Allee 11, D-89069 Ulm, Germany} \date{\today}
\begin{abstract} A chain of singly-charged particles, confined by
a harmonic potential, exhibits a sudden transition to a zigzag
configuration when the radial potential reaches a critical value,
depending on the particle number. This structural change is a
phase transition of second order, whose order parameter is the
crystal displacement from the chain axis. We study analytically
the transition using Landau theory and find full agreement with
numerical predictions by J. Schiffer [Phys. Rev. Lett. {\bf 70},
818 (1993)] and Piacente {\it et al} [Phys. Rev. B {\bf 69},
045324 (2004)]. Our theory allows us to determine analytically the
system's behaviour at the transition point. \end{abstract}

\maketitle

\section{Introduction}

Wigner crystals of ions in Paul or Penning traps are a remarkable
example of selforganized matter at ultralow
temperatures~\cite{DubinRMP}. These systems are usually composed
of singly-charged particles, which are kept together by external
time-dependent radio-frequency or static magneto-electric
potentials, and which reach crystallization by means of laser
cooling. Among several important aspects, the transition from
disorder to order for few ions was studied
in~\cite{Diedrich,Bluemel1988,ChaosOrder}; long-range order in
three dimensional structures in Penning traps was first
demonstrated in~\cite{Tan1995,Itano98}; and more complex
crystalline structures have been realized, see for
instance~\cite{Drewsen01,Drewsen03,Drewsen06}. Most recently,
these crystalline structures have been attracting increasing
attention for the realization of quantum information
processors~\cite{Cirac95,QLogic:2,QLogic:3} and
simulators~\cite{wund02,Porras,Pons07,Calarco}. In this perspective, the
clear understanding and characterization of the structural
properties would provide the possibility to control at the
microscopic level the dynamics of complex systems. Moreover, ion
crystals are systems characterized by truly long-range,
unscreened Coulomb interactions, and hence constitute interesting
physical systems where one can test equilibrium and
out-of-equilibrium statistical mechanics models for systems
exhibiting non-extensive thermodynamic
functions~\cite{RuffoBook,DubinScience00}.

\begin{figure} \includegraphics[width=1.2\columnwidth]{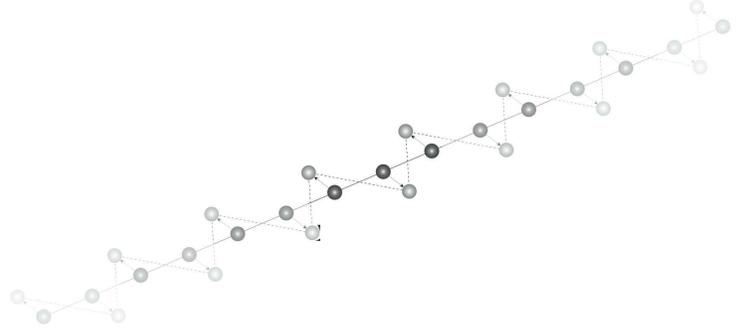}
\caption{\label{fig:1} Structural phase transition in a string of equidistant
trapped ions from a linear to a planar zigzag configuration. For
ions in a harmonic trap, close to the transition point the zigzag
configuration is evident about the center of the trap, where the
density of ion is larger~\cite{Schiffer93}.} \end{figure}

Structural transitions in ion crystals are induced either by
changing the external potential~\cite{MPQ1,MPQ2} or by introducing
other forms of instabilities~\cite{Stick-slip}. Structural
transitions in low dimensional ion crystals were first
characterized experimentally in~\cite{MPQ1,MPQ2}. Here, starting
from a chain configuration, the sudden transition to a planar
zigzag structure, as shown in Fig.~\ref{fig:1}, was observed when
the radial potential reached a critical value, dependent on the
ion number. In theoretical investigations it was conjectured that
the structural change from a chain to a zigzag is a second order
phase transition~\cite{Schiffer93}. Further numerical work showed
that at this transition point the ground state energy is
characterized by a discontinuity in the second
derivative with respect to the particles density~\cite{Piacente2005}.

%%%%%%%%%%%%%%%%%%%%%%%%%
%%%%%%%%%%%%%%%%%%%%%%%%%
In this article we study the structural phase transition of an ion
crystal from a linear chain to a zigzag configuration in a
suitably defined thermodynamic limit, by developing an analytic
theory which allows us to determine the behavior of the system at
the critical point. 
From symmetry considerations we conjecture the spontaneous symmetry breaking.
Applying Landau theory~\cite{Landau}, we identify the order
parameter and the soft mode driving the instability, and
demonstrate that the system undergoes a second order phase
transition. Our theory is valid at $T=0$, when the system exhibits long-range order. It allows us to determine the system's behavior at the transition point, and the results we find are in agreement with the
numerical results reported in~\cite{Schiffer93} and
in~\cite{Piacente2005}.

This article is organized as follows. In Sec.~\ref{Sec:Model} we
introduce the model and discuss first the transition for a chain
of 3 ions from a linear to a zigzag configuration of charges. In
Sec.~\ref{Sec:Th-Limit} we derive the dispersion relations and
eigenmodes at equilibrium of the linear chain and of the zigzag
configuration in the thermodynamic limit. In Sec.~\ref{Sec:Phase}
we focus onto the classical phase transition between the two
configurations, identify the soft mode and study analytically the
system around the critical point. In Sec.~\ref{Sec:Conclusions} we
conclude and in the appendices we report the details of
calculations presented in Sec.~\ref{Sec:Phase}.

\section{Ordered structures of ions in low dimensions}
\label{Sec:Model}

The model we consider is constituted by $N$ particles of mass $m$
and charge $Q$, which are confined by an external harmonic
potential along one axis. The particles are classical, and the
Hamiltonian governing their dynamics reads \begin{equation}
\label{H:0} H=\sum_{j=1}^N\frac{{\bf p}_j^2}{2m}+ V({\bf
r}_1,\ldots,{\bf r}_N)\;, \end{equation} where ${\bf
r}_j=(x_j,y_j,z_j)$ and ${\bf p}_j$ are the positions and
conjugate momenta, with $j=1,\ldots, N$. The term $V$ accounts for
the oscillator's potential and the Coulomb repulsion,
\begin{eqnarray} V
=\frac{1}{2}\sum_{j=1}^Nm\left[\nu^2x_j^2+\nu_t^2(y_j^2+z_j^2)\right]+\frac{1}{2}\sum_{j=1}^N\sum_{j\neq
i}\frac{Q^2}{|{\bf r}_i-{\bf r}_j|}\;. \nonumber\\
\label{Eq:potential}\end{eqnarray} Here, the potential is
characterized by harmonic confinement at frequency $\nu$ and
$\nu_t$ in the axial and transverse direction, respectively,
whereby $\nu_t> \nu$ for the case we are going to study. We denote
by $\alpha\equiv\nu_t/\nu$ the trap aspect ratio, such that
$\alpha>1$.

At sufficiently low temperatures, the ions localize themselves at
the equilibrium positions ${\bf r}_j^{(0)}$ which solve the
coupled equations describing the equilibrium of the forces,
\begin{eqnarray} \label{Eq:Discrete} \left. \frac{\partial
V}{\partial {\bf r_j}}\right|_{{\bf r_j}={\bf r_j^{(0)}}}=0\;.
\end{eqnarray} When the transverse frequency $\nu_t$ exceeds a
critical value $\nu_t^{(c)}$, which depends on the axial trap
frequency $\nu$ and on the number of ions, the solutions of
Eq.~(\ref{Eq:Discrete}) are aligned along the $x$-axis, forming a
string. Tables of the equilibrium positions for string up to 10
ions have been reported in~\cite{Steane,James98}. An analytical
form for the linear density of ions along the trap axis at
equilibrium was determined in~\cite{Dubin97} for $N\gg 1$ and
using the local density approximation. The linear fluctuations
about the classical ground state of an ion chain in a harmonic
trap have been analytically studied
in~\cite{Morigi04prl,morigi-pre}. This study identified as well
the value of the critical transverse frequency $\nu_t^{(c)}\approx
3N\nu/(4\sqrt{\log N})$, using an expansion at leading order in
$1/\log N$, and by considering only nearest-neighbour
contributions. Within this approximation this value is consistent
with numerical results~\cite{Schiffer93}, and is in good agreement
with previous analytical evaluations in~\cite{Dubin93}, which
calculated the critical value taking into account the long-range
interaction between the ions but assuming that the particles are
equidistant.

When the transverse frequency is varied, so that
$\nu_t<\nu_t^{(c)}$, the stable configuration is first a zigzag
structure, then at smaller values it has an abrupt transition to a
helicoidal one, and so on thereby acquiring more complex
structures~\cite{MPQ1,MPQ2,Dubin93}. Eventually, for a large
number of ions and for aspect ratios $\alpha$ sufficiently close
to unity the structure is expected to take the b.c.c. crystalline
form~\cite{DubinRMP}. In the following, we study the transition
from an ion chain to a zigzag structure for the most simple model,
namely three ions in a linear Paul trap. This system allows us to
get some insight into the system, before considering the
structural transition in the thermodynamic limit in
Sec.~\ref{Sec:Th-Limit}.

\subsection{Structural stability of a three-ion chain}
\label{Sec:3:ions}

We consider $N=3$ ions inside a trap with $\nu_t>\nu$, and
calculate their equilibrium positions as a function of the aspect
ratio $\alpha=\nu_t/\nu$. We restrict for simplicity to two
dimensions, which we here identify with the $x-y$ plane, and
rewrite the potential \eqref{Eq:potential} in dimensionless
variables as \begin{equation} \tilde
V=\frac{1}{2}\sum_{i=1}^3({x'}_i^2+\alpha^2 {y'}_i^2)+\sum_{i<j}
\frac{1}{\sqrt{(x'_i-x'_j)^2+ (y'_i-y'_j)^2 }}\;, \end{equation}
where $x'_i= x_i/l$, $y'_i= y_i/l$ , $l^3=Q^2/(m\nu^2)$ and
$\tilde V=V/(l^2 m \nu^2)$. Throughout this section we drop the
prime superscript.

%%%%%%%%%%%%%%%%%%%%%%%%%%
\begin{figure}
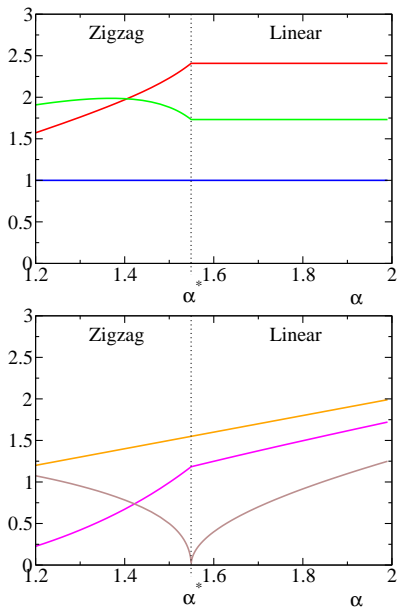

\includegraphics[width=0.6\columnwidth]{fig2}
\includegraphics[width=0.6\columnwidth]{fig3}
\caption{\label{fig:modes} (color online) Excitation frequencies
of the oscillations modes of a three-ion string, in units of the
axial trap frequency $\nu$, as a function of the trap anisotropy
$\alpha$. The eigenmodes for $\alpha>\alpha^*$ are the longitunal
(top) and the transverse ones (bottom). For $\alpha<\alpha^*$ the
eigenmodes are combination of longitudinal and transverse modes
with opposite parity by reflection about $x=0$.} \end{figure}
\begin{figure}
\includegraphics[width=0.6\columnwidth]{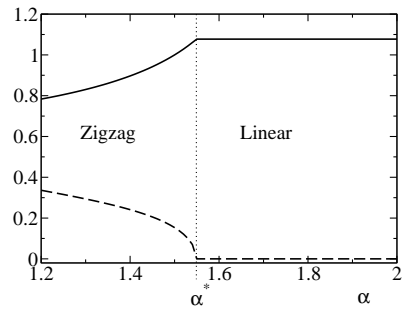}
\caption{\label{fig:solutions} Equilibrium position of the
external ions of a string of 3 particles as a function of the trap
anisotropy $\alpha$. The solid and dashed lines display the
longitudinal and transverse variables, $\bar x$ and $\bar y$, in
units of the characteristic length $l$. The vertical dotted line
indicates the transition value $\alpha^*$, where the equilibrium
configuration makes an abrupt change from a linear chain to a
zigzag structure. } \end{figure}

%%%%%%%%%%%%%%%%%%%%%%%%%%
The normal modes frequencies for the linear and the zigzag
structures are displayed in Fig.~\ref{fig:modes} as a function of
$\alpha$, when $\alpha$ is decreased across the value for which
the linear chain becomes mechanically unstable. The transition to
the zigzag configuration takes place at the value
$\alpha=\alpha^*$, such that the smaller transverse frequency of
the linear chain vanishes. We now study in detail the classical
equilibrium positions for $\alpha>1$. Assuming the convention
$x_1<x_2<x_3$, the symmetry of the trapping potential imposes
$x_2^{(0)}=0$, $x_1^{(0)}=-x_3^{(0)}\equiv-\bar{x}$, with
$\bar{x}>0$. From Eqs.~(\ref{Eq:Discrete}) we also find
$y_2^{(0)}=-2\bar y$ and $y_1^{(0)}=y_3^{(0)}\equiv\bar y$, with
$\bar y\ge 0$. The linear configuration, namely the set of
solutions with $\bar y=0$, is found when the aspect ratio
$\alpha>\alpha^*$, where $\alpha^*\equiv\sqrt{12/5}$. When
$\alpha<\alpha^*$, then $\bar y>0$ and the structure becomes
planar. We denote this case by "zigzag configuration", as it is
indeed the most elementary instance of the structure one observes
for many ions. Here, for $\alpha<\alpha^*$ terms $\bar x$, $\bar
y$ take the form \begin{eqnarray} \bar
x=\left[4\left(1-\frac{\alpha^2}{3}\right)\right]^{-\frac{1}{3}},~~~
\bar y=
\frac{1}{3}\sqrt{\left(\frac{3}{\alpha^2}\right)^{\frac{2}{3}}-\bar
x^2}, \label{ybsolution} \end{eqnarray} Their functional
dependence on the aspect ratio $\alpha$ is displayed in
Fig.~\ref{fig:solutions}. One can observe the discontinuity of the
derivative at $\alpha^*$, corresponding to the transition to a
different equilibrium configuration. For $\alpha\to\alpha^*$ the
change is faster for the transverse displacement, as it is visible
by the expansion of $\bar x,\bar y$ at
$\delta\alpha=\alpha^*-\alpha$, \begin{eqnarray*} \bar
y&=&y_0~\delta\alpha^{\frac{1}{2}}+{\rm
O}\left(\delta\alpha^{3/2}\right)\\
\bar x&=&\bar x_{\rm lin}-x_0~\delta\alpha+ {\rm
O}\left(\delta\alpha^{2}\right) \end{eqnarray*} where $\bar x_{\rm
lin}=(5/4)^{1/3}$ is the value  taken by $\bar x$ when the linear
chain is stable, while $y_0\approx 0.74$, $x_0\approx 1.85$.
%%%%%%%%%%%%%%%%%%%%%%%%%
%%%%%%%%%%%%%%%%%%%%%%%%%
We note that, about the instability point of the linear chain, the
transverse displacement $\bar y$ plays the role of the order
parameter, while the changes of the axial distance $\bar x$ are
induced by the changes of $\bar y$, and therefore about the value
$\alpha^*$ these are less dramatic.

\section{The linear and the zigzag structures} \label{Sec:Th-Limit}

In this section, we study the static properties of the linear
chain and of the zigzag configuration in the thermodynamic limit.
For an ion chain inside a trap, a good thermodynamic limit is
found by fixing the interparticle spacing $a$ at the chain center
when $N\to \infty$. This corresponds to the requirement that the
axial trap frequency vanishes according to the relation $\nu\sim
\sqrt{\log N}/N$~\cite{Morigi04prl,morigi-pre}. In this limit, the
critical transverse frequency $\nu_t^{(c)}$ is constant, and the
behaviour at the mechanical instability is equivalent to that of a
uniform chain with equal interparticle distance $a$ between
neighbouring ions~\cite{MorigiJPB}. The uniform chain is the model
we will use for determining the ground state and the motion of the
linear and zigzag structure in the thermodynamic limit.

\subsection{The linear chain} \label{Sec:Linear_Chain}

We assume a stable linear chain of ions, namely $\nu_t
>\nu_t^{(c)}$. In this limit the equilibrium positions lie along the
$x$-axis, ${\bf r}_j^{(0)}=(x_j^{(0)},0,0)$, and we use the
convention $x_i>x_j$ for $i>j$. For small vibrations around these
points we approximate the potential in Eq.~(\ref{Eq:potential}) by
its second order Taylor expansion in the displacements
$q_j=x_j-x_j^{(0)}$, $y_j$, $z_j$. In this limit the equations of
motion are 
\begin{eqnarray} \label{Eq:ax}
&&\ddot{q}_i=-\nu^2q_i-\sum_{j\neq i}\frac{\mathcal K_{i,j}}{m}(q_i-q_j)\;,\\
&&\ddot{y}_i=-\nu_t^2y_i+\frac{1}{2}\sum_{j\neq i}\frac{\mathcal
K_{i,j}}{m}(y_i-y_j)\;,
\label{Eq:y}\\
&&\ddot{z}_i=-\nu_t^2z_i+\frac{1}{2}\sum_{j\neq i}\frac{\mathcal
K_{i,j}}{m}(z_i-z_j)\;, \label{Eq:z} \end{eqnarray} and describe a
system of coupled oscillators, with long range interaction and
position-dependent coupling strength. Here, the coefficients
$\mathcal K_{i,j}\equiv-\partial^2 V/\partial x_j\partial
x_i|_{x_j^{0}}$ read \begin{equation} \label{eq:kij}
\mathcal
K_{i,j}=\frac{2Q^2}{|x_i^{(0)}-x_j^{(0)}|^3}\;. \end{equation} We
note that at second order in the harmonic expansion the axial and
transverse vibrations are decoupled. It is easily verified that
the center-of-mass motion is an eigenmode of the secular
equations~(\ref{Eq:ax})-(\ref{Eq:z}) at eigenfrequencies $\nu$ and
$\nu_t$ for the axial and transverse motion, respectively. The
solution to Eqs.~(\ref{Eq:ax})-(\ref{Eq:z}) have been studied
in~\cite{Morigi04prl,morigi-pre}.
%%%%%%%%%%%%%%%%%%%%%%%%
\begin{figure}
\rput{0}(-0.5,2.5){$\displaystyle{\frac{\omega}{\omega_0}}$}
\includegraphics[scale=0.2]{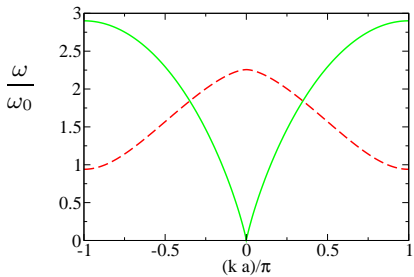}
\caption{\label{fig:spectrum-lin}(color online) Excitation
spectrum of the uniform chain.  The eigenfrequencies $\omega$, in
units of $\omega_0=\sqrt{Q^2/ma^3}$, are plotted as a function of
the quasimomentum $k$, in units of $\pi/a$. The axial spectrum
(green solid line) and the transverse spectrum (red dashed line) are obtained
from Eqs.~(\ref{eq:linspectrum1}) and~(\ref{eq:linspectrum2}),
respectively. Here, $\nu_t=1.1\nu_t^{(c)}$. } \end{figure}
%%%%%%%%%%%%%%%%%%%%%%%%%

For the purpose of studying the behaviour at the mechanical
instability, we now consider the simplified model of the uniform
chain, where the interparticle distance at equilibrium is fixed.
This case is found by setting $\nu=0$ in
Eqs.~(\ref{Eq:ax})-(\ref{Eq:z}) and assuming constant interparticle distance
$a=x_{j+1}^{(0)}-x_{j}^{(0)}$. Such condition can be realized
for the central ions of a long ion chain inside of a linear Paul trap \cite{raizen92} or for ions confined in a ring of large radius\cite{MPQ1,MPQ2}. This second scenario corresponds to
take periodic boundary conditions, $q_1=q_{N+1}$, etc.
Crystallization is found assuming, for instance, that one ion is
pinned at the position $x_0^{(0)}=0$. Then, the classical
equilibrium positions are $x_j^{(0)}=ja$ and the coupling
strengths in Eqs.~(\ref{Eq:ax})-(\ref{Eq:z}) take the form
\begin{equation} \label{Ki-j} \mathcal K_{i,j}^{\rm
uniform}=\frac{2Q^2}{|i-j|^3a^3}\equiv\mathcal K_{i-j}.
\end{equation} The  dispersion relations
are~\cite{Ashcroft}
\begin{eqnarray}
\omega_{\|}(k)^2= 4\left(\frac{2Q^2}{ma^3}\right)\sum_{j=1}^N\frac{1}{j^3}\sin^2\frac{jka}{2}\;,
\label{eq:linspectrum1}\\
\omega_{\perp}(k)^2=
\nu_t^2-2\left(\frac{2Q^2}{ma^3}\right)\sum_{j=1}^N\frac{1}{j^3}\sin^2\frac{jka}{2}\;,
\label{eq:linspectrum2} 
\end{eqnarray} 
with $k=2\pi n/Na$ and
$n=0,\pm 1,\pm 2,\ldots,N/2$. The spectrum corresponding to
Eqs.~\eqref{eq:linspectrum1} and~\eqref{eq:linspectrum2} is shown
in Fig.~\ref{fig:spectrum-lin}. The axial eigenmodes at frequency
$\omega_{\|}(k)$ are $\Theta_{k}^{(\pm)}$, such that
\begin{equation} \label{Fourier:q} q_j=\sqrt{\frac{2}{N}}\sum_{k>0}\left(\Theta_k^{(+)}\cos kja+\Theta_k^{(-)}\sin
kja\right)\;, \end{equation} where the superscript $\pm$ indicates
 parity by reflection $k\to -k$. Analogously, we denote the
transverse eigenmodes at frequency $\omega_{\perp}(k)$ by
$\Psi_k^{y(\pm)}$ and $\Psi_k^{z(\pm)}$, where \begin{eqnarray}
\label{Fourier:y}
y_j&=&\sqrt{\frac{2}{N}}\sum_{k>0}\left(\Psi_k^{y(+)}\cos kja+\Psi_k^{y(-)}\sin kja\right)\;,\\
z_j&=&\sqrt{\frac{2}{N}}\sum_{k>0}\left(\Psi_k^{z(+)}\cos
kja+\Psi_k^{z(-)}\sin kja\right)\,. \label{Fourier:z}
\end{eqnarray}
We note that the modes at $k=\pi/2$ are even. 
A close inspection to Eq.~(\ref{eq:linspectrum2}) shows that there
may exist values of the transverse trap frequency, at fixed
interparticle distance $a$, for which $\omega_{\perp}^2<0$, that
is, imaginary frequency solutions. For such values, thus, the
chain is unstable. The threshold value $\nu_t^{(c)}$, such that
for $\nu_t>\nu_t^{(c)}$ the linear chain is stable, is found by
solving ${\rm min}_k(\omega_{\perp})=0$ (see
Sec.~\ref{Sec:soft:mode}). The minimum is found at $k=\pi/a$  and
correspondingly \begin{eqnarray} \nu_t^{(c)~2}&=&
2\left(\frac{2Q^2}{ma^3}\right)\sum_{j=1}^N\frac{1}{j^3}\sin^2\frac{j~\pi}{2}\nonumber\\
&\to&\frac{Q^2}{ma^3}\frac{7}{2}\zeta(3)\;, \label{nu:t:c}
\end{eqnarray} where result~(\ref{nu:t:c}) is found for $N\to
\infty$ using
$\sum_{\ell>0}(2\ell-1)^{-p}=\left(1-2^{-p}\right)\zeta(p)$, with
$\zeta(p)$ the Riemann-zeta function. The value in
Eq.~(\ref{nu:t:c}) depends on the interparticle spacing $a$ and
provides the range of validity of the results presented in this
section. It coincides with the value reported in~\cite{Dubin93},
where a similar model to the one discussed here was considered. It
is close to the result $\nu_t^{(c),{\rm trap}2}=4Q^2/ma_{\rm
trap}(0)^3$ found at leading order in $1/\log N$
in~\cite{morigi-pre}, where $a(x)$ gives the interparticle
distance as a function of $x$ in the local density approximation,
and $a(0)\equiv a$ is the value at the chain center. This result
was obtained by considering the inhomogeneous distribution of ions
along the chain, but keeping only the nearest-neighbours
interaction. The small discrepancy between the two values is to be
attributed to the different approximations that have been applied
in each model.

\subsection{The zigzag structure}

\begin{figure}
\rput{0}(3.5,-0.4){$\displaystyle{\nu_t/\nu_t^{(c)}}$}
\includegraphics[scale=0.2]{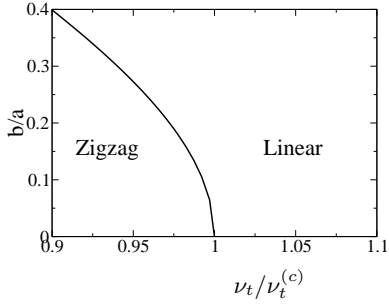} \vspace{0.8cm}
\caption{\label{fig:eqpos-therm} Transverse equilibrium
displacement $b$, in units of the interparticle spacing $a$, as a
function of the transverse frequency $\nu_t$ in units of
$\nu_t^{(c)}$. On the right of the curve the ion crystal is a
linear chain. In the region on the left of the curve it exhibits a
zigzag structure.} \end{figure}

For $\nu_t<\nu_t^{(c)}$, and sufficiently close to the critical
value, the stable configuration is a zigzag structure. We now
evaluate its dispersion relation and eigenmodes for ions on a ring
and for periodic boundary conditions. We assume the equilibrium
positions to lie on the $x-y$ plane with ${\bf
r}_n^{(0)}=(x_n^{(0)},y_n^{(0)},0)$. Then, $x_n^{(0)}=na$ and
$y_n^{(0)}=(-1)^nb/2$, with $b$ a real and positive constant,
which is determined from the equation \begin{eqnarray}
\label{Eq:b}
\frac{m\nu_t^2}{Q^2}-\sum_{\ell>0}\frac{4}{[(2\ell-1)^2a^2+b^2]^{3/2}}=0\;.
\end{eqnarray} Figure~\ref{fig:eqpos-therm} displays the
transverse equilibrium displacement $b$ as a function of the
transverse frequency $\nu_t$, as it is obtained by solving
numerically Eq.~(\ref{Eq:b}).
%%%%%%%%%%%%%%%%%%%%%%%%%
%%%%%%%%%%%%%%%%%%%%%%%%%
Assuming that the zigzag configuration is stable, we denote by
$q_n=x_n-x_n^{(0)}$, $w_n=y_n-y_n^{(0)}$, and $z_n$ the axial and
transverse displacements, and expand the potential of
Eq.~(\ref{Eq:potential}) up to second order. In this limit the
motion along the $z$ direction is decoupled from the vibrations on
the plane, and the resulting equations of motion for $q_n$ and
$w_n$ read \begin{eqnarray} \label{Eq:ax:zz}
m\ddot{q_n}&=&-\sum_{\ell \neq 0}\mathcal K^x_\ell(q_n-q_{n+\ell})\\
& &-(-1)^n\sum_{\ell \neq 0}\mathcal Y_\ell(w_n-w_{n+\ell})\;, \nonumber\\
m\ddot{w_n}&=&-m\nu_t^2w_n+\sum_{\ell\neq 0}\mathcal
K_\ell^y(w_n-w_{n+\ell})\label{Eq:y:zz}\\
& &-(-1)^n\sum_{\ell \neq 0}\mathcal
Y_\ell(q_n-q_{n+\ell})\;.\nonumber \end{eqnarray} The coefficients
appearing in these equations depend only on the interparticle
distance, $\ell a=(n'-n)a$, as the structure is periodic along
$x$. In particular, for $\ell$ even they read \begin{eqnarray*}
{\cal K}_{\ell}^x=2{\cal
K}_{\ell}^y=\frac{2Q^2}{a^3}\frac{1}{|\ell|^3}\;,~~~{\cal
Y}_{\ell}=0\;, \end{eqnarray*} while for $\ell$ odd they are given
by \begin{eqnarray*}
&&{\cal K}_{\ell}^x=\frac{Q^2}{a^3}\frac{2\ell^2-\chi^2}{[\ell^2+\chi^2]^{5/2}}\;,\\
&&{\cal K}_{\ell}^y=\frac{Q^2}{a^3}\frac{\ell^2-2\chi^2}{[\ell^2+\chi^2]^{5/2}}\;,\\
&&{\cal Y}_{\ell}=\frac{Q^2}{a^3}\frac{3\ell
\chi}{[\ell^2+\chi^2]^{5/2}}\;, \end{eqnarray*} with $\chi=b/a$.
The coefficients Eq.~(\ref{Ki-j}), and the corresponding equations
of motion for the linear chain, Eqs.~(\ref{Eq:ax}),~(\ref{Eq:y}),
are recovered for $\chi\to 0$.

\begin{figure}
\rput{0}(-0.5,2.5){$\displaystyle{\frac{\omega}{\omega_0}}$}
\includegraphics[scale=0.2]{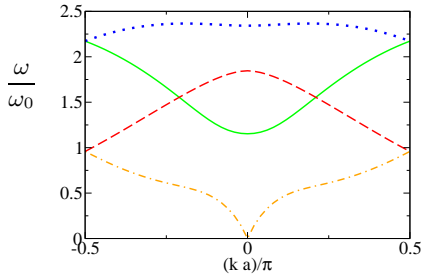}
\caption{\label{fig:spectrum-zz} (color online) Branches of the
excitation spectrum of a zigzag structure for the modes on the
$x-y$ plane, as obtained from Eq.~(\ref{om:ax:zz}). The curves
display the frequencies $\omega_{2,+}(k)$ (green solid),
$\omega_{2,-}(k)$ (blue dotted), $\omega_{1,+}(k)$ (orange dot-dashed),
$\omega_{1,-}(k)$ (red dashed), in units of $\omega_0$, as a function
of $k$, in units of $\pi/a$. The Brillouin zone is now half the
Brillouin zone of the linear chain due to the doubling of the
crystal periodicity. Here, $\nu_t=0.9\nu_t^{(c)}$.} \end{figure}

In general, the structural change brings to a doubling of the unit
cell $d$ of the crystal, which from $d=a$ in the linear chain goes
to $d=2a$ in the zigzag configuration. Correspondingly, the
Brillouin zone of the zigzag is reduced by a factor 2, and the
wave vectors now take the values
$k=2\pi n/Na$ and
$n=0,\pm 1,\pm 2,\ldots,N/4$. In
Eqs.~(\ref{Eq:ax:zz}) and~(\ref{Eq:y:zz}) one can easily verify
that the bulk excitations are eigenmodes of the chain at
frequencies $\nu$ and $\nu_t$. The other eigenvalues and
eigenfunctions can be found using the ansatz ${\bf
f}_n^{(j,\pm)}$, with 
\begin{eqnarray} 
{\bf f}_n^{(j,\pm)}(k)=(\pm
1)^n{\rm e}^{-i\omega_{j,\pm} t+ikna}\left[{\bf \hat{x}}
\mp ie^{-in\pi}\epsilon_{k}^{(j,\pm)}{\bf \hat{y}}\right]\;,
\end{eqnarray} where $j=1,2$ and $ka$ varies on the interval
$[-\pi/2,\pi/2]$. In particular, we note the relation ${\bf
f}_n^{(j,-)}(k)={\bf f}_n^{(j,+)}(k+\pi/a)$. The corresponding
eigenmodes are given by the real and imaginary parts of these
vectors. Using this ansatz, we obtain the coupled equations
\begin{eqnarray} 
&&\omega_{j,\pm}(k)^2
=C_1^{(\pm)}(k)+\epsilon_k^{(j,\pm)}B(k)\;,\label{eps:1}\\
&&(\nu_t^2-\omega_{j,\pm}(k)^2)=C_2^{(\pm)}(k)-{\epsilon_k^{(j,\pm)}}^{-1}B(k)\;,
\label{eps:2}\end{eqnarray}  
whereby 
\begin{eqnarray*}
&&B(k)=\frac{2}{m}\sum_{\ell > 0} \mathcal Y_{2\ell-1}\sin(2\ell-1)ka\;,\\
&&C_1^{(+)}(k)=\frac{4}{m}\sum_{\ell>0}\mathcal K_\ell^x\sin^2\frac{k\ell a}{2}\;, \\
&&C_2^{(+)}(k)=\frac{4}{m}\sum_{\ell>0}\left(\mathcal
K_{2\ell}^y\sin^2k\ell a+\mathcal
K_{2\ell-1}^y\cos^2\frac{(2\ell-1)ka}{2}\right)\;, \\
&&C_1^{(-)}(k)=\frac{4}{m}\sum_{\ell>0}\left(\mathcal
K_{2\ell}^x\sin^2k\ell a+\mathcal
K_{2\ell-1}^x\cos^2\frac{(2\ell-1)ka}{2}\right)\;, \\
&&C_2^{(-)}(k)=\frac{4}{m}\sum_{\ell>0}\mathcal K_\ell^y\sin^2\frac{k\ell a}{2}\;.
\end{eqnarray*}
 The eigenfrequencies are found
by eliminating the parameter $\epsilon_k^{(j,\pm)}$ from
Eqs.~(\ref{eps:1})-(\ref{eps:2}). The excitation spectrum exhibits
four branches in the new Brillouin zone, and their functional
dependence on $k$ is \begin{widetext}\begin{eqnarray}
\label{om:ax:zz}
&&\omega_{j,\pm}(k)^2=\frac{\nu_t^2+C_1^{(\pm)}(k)-C_2^{(\pm)}(k)}{2}+(-1)^j\sqrt{\frac{(\nu_t^2-C_1^{(\pm)}(k)-C_2^{(\pm)}(k))^2}{4}
+B(k)^2}\;,
\end{eqnarray}\end{widetext} with $j=1,2$. The spectrum for the
excitations on the $x-y$ plane is displayed in
Fig.~\ref{fig:spectrum-zz}.

%%%%%%%%%%%%%%%%%%%%%%%%
%%%%%%%%%%%%%%%%%%%%%%%%%
We note that in the limit $b\to 0$ the branches of the spectrum of
the linear chain, Eqs.~(\ref{eq:linspectrum1})
and~(\ref{eq:linspectrum2}), are recovered from
Eqs.~(\ref{om:ax:zz}). In fact, for $b=0$ we have $B=0$
and $C_{1}^{(\pm)}=2C_{2}^{(\mp)}$, such that each solution has
double degeneracy, with
$$\omega_{2,+}(k)^2\Bigl|_{b=0}=\omega_{2,-}(k)^2\Bigl|_{b=0}=C_1^{(+)}(k)\Bigl|_{b=0}$$
and
$$\omega_{1,+}(k)^2=\omega_{1,-}(k)^2=\nu_t^2-C_1^{(+)}(k)\Bigl|_{b=0}/2,$$
which reproduce respectively Eqs.~(\ref{eq:linspectrum1})
and~(\ref{eq:linspectrum2}) (note that $\nu_t^2-C_1^{(\pm)}(k)-C_2^{(\pm)}(k)<0$).

\section{Landau theory of the structural phase transition}
\label{Sec:Phase}

If the ions are crystallized along a line, by lowering the
transverse confinement $\nu_t$ the system will be led to a
situation in which the linear chain gets unstable. In this
regime, one observes experimentally a transition, in which the ions are
crystallized on a plane, according to a zigzag distribution of
particles. In the literature it was conjectured that this is a
second-order phase transition. This conjecture is supported by the
numerical results in~\cite{Schiffer93,Piacente2005}.

Indeed, one can observe that the transition from a linear to a
zigzag configuration is characterized by a symmetry breaking
resulting in the increase of the unit cell by a factor of 2. It is
combined with a transition from a linear to a planar structure
corresponding to the loss of rotational symmetry about the $x$-axis.
Then, one can identify the order parameter with the displacement
of the equilibrium position from the $x$-axis, while the control
parameter can be taken as the transverse frequency $\nu_t$ when
the interparticle distance is fixed. Starting from this educated
guess we apply Landau theory to the transition~\cite{Landau}. We
focus on the situation in which the interparticle distance $a$ is
fixed, and study the crystal structure when the transverse
confinement $\nu_t$ varies across the critical value
$\nu_t^{(c)}$. We explicitly determine the critical exponent of
the order parameter around the critical value, and find that it is
in agreement with the numerical results in~\cite{Schiffer93}.

%%%%%%%%%%%%%%%%%%%%%%%%%

\subsection{The soft mode} \label{Sec:soft:mode}

\begin{figure}
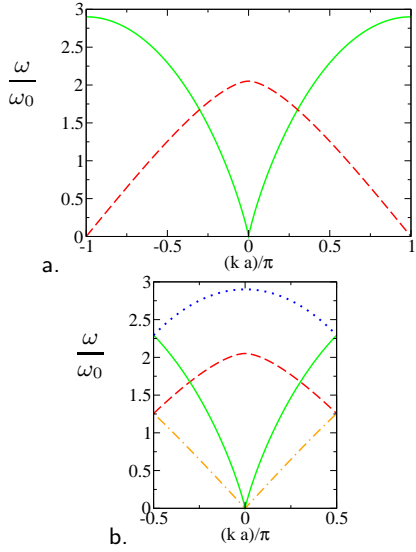
 {\sf
a.}\rput{0}(-0.5,2.5){$\displaystyle{\frac{\omega}{\omega_0}}$}
\begin{centering}\includegraphics[scale=0.2]{fig8}\end{centering}\\
{\sf
b.}\rput{0}(-0.5,2.5){$\displaystyle{\frac{\omega}{\omega_0}}$}
\begin{centering}\includegraphics[scale=0.2]{fig9}\end{centering}\\
\caption{ \label{fig:spectrum-pt} (color online) Branches of the
excitation spectrum at ({\sf a.}) $\nu_t=\nu_t^c+0^+$ (just above
the critical value) and ({\sf b.}) $\nu_t=\nu_t^c+0^-$ (just below
the critical value). In ({\sf b.}) the equilibrium structure is a
zigzag, the periodicity is doubled with respect to the linear
chain and the new Brillouin zone is halfed. The four branches of
the spectrum are obtained at this point by ``folding'' the two
branches of the linear chain in ({\sf a.}). The units and style
codings in ({\sf a.}) and ({\sf b.}) are the same as in
Figs.~\ref{fig:spectrum-lin} and~\ref{fig:spectrum-zz},
respectively. Here, $\nu_t^{(c)}\simeq 2.05\omega_0$.}
\end{figure}

Let us now go back to the dispersion relation for the transverse
modes of the linear chain in Eq.~(\ref{eq:linspectrum2}). The
structural transition takes place for the critical value
$\nu_t^{(c)}$, Eq.~(\ref{nu:t:c}), such that the frequency of the
lowest transverse mode of the linear chain vanishes, as shown in
Fig.~\ref{fig:spectrum-pt}a. The smallest transverse frequency
$\omega_{\perp}$ is found at the value of the wave vector $k$, at
which the semipositive-definite function \begin{eqnarray*}
F(\varphi)=\sum_{j=1}^N\frac{1}{j^3}\sin^2(j\varphi)
\end{eqnarray*} is maximum in the interval $0\le\varphi\le\pi/2$,
as seen from Eq.~(\ref{eq:linspectrum2}) for $ka=[0,\pi]$. We
first observe that $\partial F/\partial \varphi=0$ at
$\varphi=0,\pi/2$. As $F(0)=0$, at $\varphi=0$ the function has an
absolute minimum. The second-order derivative at $\varphi=\pi/2$
is negative, and one can simply prove analytically that this point
is at least a relative maximum. Numerical studies show that it is
an absolute maximum, such that the smallest transverse frequency
is found at wave vector $k_0\equiv\pi/a$ and takes the value
\begin{equation} \omega_{\perp,{\rm min}}^2=
\nu_t^2-\nu_t^{(c)~2}.\label{w:perp:min}\end{equation} This
identifies the soft mode. The corresponding eigenmodes exhibit a
periodic deformation of the chain at periodicity $2a$, analogous
to the zigzag structure. We denote by $b_0$ the amplitude of its
oscillations, with $b_0=b_0(\nu_t)$, such that the transverse
oscillations along $y$ of the ion $j$ are described by the
function \begin{equation} \label{soft:mode} y^{\rm soft}_j=(-1)^j
b_0/2. \end{equation} In the following we assume zero temperature
and study the equilibrium position of the crystal with the
transverse frequency varying in the interval
$[\nu_t^{(c)}-\delta\nu,\nu_t^{(c)}+\delta\nu]$, thus on both
sides of the critical point, whereby $\delta\nu$ is a small
positive quantity. Following Landau theory, we demonstrate that
the zigzag mode of the linear chain, given by
Eq.~(\ref{soft:mode}), is indeed the soft mode, driving the
instability across the critical point, and we evaluate the
critical exponents for some quantities of interest.

%%%%%%%%%%%%%%%%%%%%%%%%
%%%%%%%%%%%%%%%%%%%%%%%%

\subsection{Equilibrium positions around the critical point}
\label{sec:soft}

In order to determine the behaviour at the critical point, we
first expand Eq.~(\ref{Eq:potential}) till the fourth order around
the equilibrium positions of the chain, $V=\sum_{l=1}^4V^{(l)},$
where $l$ labels the order. The zero order term at leading order
in $1/N$ is~\cite{Dubin97} \begin{eqnarray} V^{(0)} =
\frac{Q^2}{a}(N-1)\left(\gamma-\ln 2+\ln(N)+{\rm
O}\left(\frac{1}{N^2}\right)\right)\;,\nonumber\\ \end{eqnarray}
where $\gamma=0.577216\dots$ is Euler's constant. The first order
term vanishes as a result of the requirement that we are looking
for a minimum. Using the decomposition into the eigenmodes of the
linear chain, Eqs.~(\ref{Fourier:q})-(\ref{Fourier:z}), the
quadratic term of the expansion of potential~(\ref{Eq:potential})
(with the use of (\ref{Eq:ax}-\ref{eq:kij}) and (\ref{eq:linspectrum1}-\ref{eq:linspectrum2}))
takes the form 
\begin{equation} \label{V:2} V^{(2)} =
\frac{m}{2}\sum_{k>0,s=\pm}
\left(\omega_{\|}(k)^2\Theta_k^{(s)2}+\beta(k)(\Psi^{y(s)2}_k+\Psi^{z(s)2}_k)\right)\,,
\end{equation} where $\omega_{\|}$ is given by
Eq.~(\ref{eq:linspectrum1}), while \begin{equation}
\label{alpha:k}
\beta(k)=\nu_t^2-2\left(\frac{2Q^2}{ma^3}\right)\sum_{j=1}^N\frac{1}{j^3}\sin^2\frac{jka}{2}
\end{equation} and it coincides with $\omega^2_{\perp}(k)$,
Eq.~(\ref{eq:linspectrum2}), for $\nu_t>\nu_t^{(c)}$. The third
and fourth order terms, obtained by using this decomposition, are
presented in App.~\ref{app:3d}.

The linear chain becomes mechanically unstable when, by varying
$\nu_t$, the frequency of the mode with wave vector $k_0=\pi/a$,
Eq.~(\ref{soft:mode}), vanishes. Starting from this observation,
we study the behaviour of the corresponding mode close to the
instability point, when $\nu_t\simeq \nu_t^{(c)}$. For
convenience, we denote by $\Psi_0^y$ and $\Psi_0^z$ the zigzag
modes of the linear chain along the $y$ and $z$ direction,
respectively, at wave vector $k_0$. Around the instability point
these modes will be coupled significantly to other
quasi-degenerate modes by the third and fourth order terms
$V^{(3)}$ and $V^{(4)}$. These quasi-degenerate modes are long
wavelengths axial modes $\Theta_{\delta k}$ at wave vectors
$\delta k$, such that $|\delta k|a\ll 1$, and short wavelength
transverse modes $\Psi^\sigma_{k_0+\delta k'}$ at wave vector
$k=k_0+\delta k'$, with $|\delta k'|a\ll 1$.

At first order in the small parameter $|\delta k|a\ll 1$, the part
$V^{(3)}_{k_0}$, that contains the summands of the third order
term $V^{(3)}$ giving the coupling of the mode at $k_0$ with the
other quasi degenerate modes, has the form \begin{widetext}
\begin{eqnarray} \label{V:3:soft}
V^{(3)}_{k_0}&=&\frac{21}{2\sqrt{2}}\zeta(3)\frac{Q^2}{a^3 \sqrt
N}~\sum_{\delta k>0}\delta k\sum_{\sigma=y,z}\Psi_{0}^\sigma
\left(\Theta^{(+)}_{\delta k}\Psi^{\sigma(-)}_{\delta
k}+\Theta^{(-)}_{\delta k}\Psi^{\sigma(+)}_{\delta
k}\right)+O(\delta k^2a^2)\;, \end{eqnarray} where we adopted for
convenience the notation 
$\Psi^\sigma_{\delta k} :=   \Psi^\sigma_{k=k_0+\delta k} $.
We note that Eq.~(\ref{V:3:soft}) is of
first order in $\delta k$. The part $V^{(4)}_{k_0}$ of the fourth
order term $V^{(4)}$, which is relevant to the dynamics of the
soft mode at $k_0$, involves only the transverse modes that are
close in $k$ to $k_0$, and has the form \begin{eqnarray}
V^{(4)}_{k_0}&=&A({\Psi^{y}_{0}}^2+{\Psi^{z}_{0}}^2)^2+12A\sum_{\sigma=y,z}{\Psi_0^{\sigma}}^2
\sum_{\delta k>0} \left[{\Psi^{\sigma(+)}_{\delta
k}}^2+{\Psi^{\sigma(-)}_{\delta
k}}^2\right]+4A\sum_{\sigma=y,z;\sigma'\neq
\sigma}{\Psi_0^{\sigma}}^2 \sum_{\delta k>0}
\left[{\Psi^{\sigma'(+)}_{\delta k}}^2+{\Psi^{\sigma'(-)}_{\delta
k}}^2\right]
\nonumber\\
&&+16A\Psi_0^{y}\Psi_0^{z}\sum_{\delta k>0}
\left[\Psi^{y(+)}_{\delta k}\Psi^{z(+)}_{\delta
k}+\Psi^{y(-)}_{\delta k}\Psi^{z(-)}_{\delta k}\right]
+A(\Psi_0^{y}+\Psi_0^{z}){\mathcal
F}\left(\Psi^{\sigma(+)}_{\delta k_1},\Psi^{\sigma'(-)}_{\delta
k_2},\Psi^{\sigma''(-)}_{\delta k_1+\delta k_2}\right)+{\rm
O}(\delta k^2 a^2)\;,\label{V:4:soft} \end{eqnarray}
\end{widetext} where $A$ is calculated from the coefficients of
$V^{(4)}$ at $k_j=k_0$ ($j=1,2,3,4$), see App.~\ref{app:3d}, and
it takes the form $A={\mathcal A}/N$ with
\begin{equation}\label{eq:A} {\mathcal A}=\frac{3}{2}\frac{31}{32}\zeta(5)\frac{Q^2}{
a^5}\;. \end{equation} The function
${\mathcal F}$ in Eq.~(\ref{V:4:soft}) contains a sum of products
of three amplitudes $\Psi_{\delta k}^{\sigma(\pm)}$ for $\delta k
\neq 0$, and it is of no importance for the following
considerations. 
The numerical factors appearing in Eq.~(\ref{V:4:soft}), multiplying each term of the sum, account for all possible permutations of the amplitudes $\Psi^\sigma_{\delta k}$ in each summand (see App.~\ref{app:3d} and
Eq.~(\ref{eq:V:4}) therein).
The coupling between the transverse modes at $k_0$ and the axial
modes does not appear explicitly in Eq.~(\ref{V:4:soft}), as it
scales with $(\delta k\; a)^2\ll 1$, and it is hence of higher
order with respect to the coupling among the transverse modes. Since the third order term, Eq.~(\ref{V:3:soft}),
scales with $\delta k\; a$, at zeroth order in the
expansion in $|\delta k|a$ and close to the instability, the
effective potential describing the dynamics of the mode at $k_0$
is given by (see Eqs.~(\ref{V:2})
and~(\ref{V:4:soft}))\begin{eqnarray} \label{V:eff}
&&V_{\rm eff}=\frac{m}{2}\beta_{0}\left[{\Psi^{y}_{0}}^2+{\Psi^{z}_{0}}^2\right]\\
&&+\frac{m}{2}\sum_{\delta k> 0}\beta_{\delta
k}\sum_{\sigma=y,z}\left[{\Psi^{\sigma(+)}_{\delta
k}}^2+{\Psi^{\sigma(-)}_{\delta
k}}^2\right]+V^{(4)}_{0}\;,\nonumber \end{eqnarray} where
$\beta_{\delta k}\equiv\beta(k_0-\delta k)$. We now allow the
transverse frequency $\nu_t$ to take values in the interval
$[\nu_t^{(c)}-\delta\nu,\nu_t^{(c)}+\delta\nu]$, such that
$\beta_{\delta k}$ may take on small but negative values. We first
determine the amplitude of the zigzag mode $k_0$ and then show
that in the vicinity of the frequency $\nu_t^{(c)}$ no other modes
are stable. For this purpose for $\beta_{\delta k}<0$ we determine
the corrections $\bar{\Psi}_{\delta k}^{y(\pm)}$,
$\bar{\Psi}_{\delta k}^{z(\pm)}$ to the equilibrium positions of
the linear chain using Eq.~(\ref{V:eff}), assuming that these give
rise to a small displacement $b$ with respect to the equilibrium
interparticle distance $a$, $b\ll a$. In particular, following our
hypothesis that close to the transition point the soft mode is
unique, and it is the zigzag mode, we consider the set of
solutions where $\bar\Psi^{(\pm)}_{\delta k}=0$ for $\delta k>0$,
and introduce the Fourier amplitude of the displacement in the
transverse plane $\bar\varrho=\sqrt{N}b/2$, as indicated from
Eq.~(\ref{Fourier:q}), such that
$$\bar{\varrho}=\sqrt{\left(\bar{\Psi}_0^{y}\right)^2+\left(\bar{\Psi}_0^{z}\right)^2}.$$
From Eq.~(\ref{V:eff}) one finds $\bar{\varrho}=0$ for
$\beta_0>0$, while for $\beta_0<0$ \begin{eqnarray}
\bar\varrho=\left(-N\frac{m\beta_0}{4{\mathcal A}}\right)^{1/2}.
\label{eq:solrho} \end{eqnarray} This is indeed a minimum if we
ignore terms in $V_{\rm eff}$ with non-zero $\delta k$. It will be
shown in what follows that this minimum is stable with respect to
addition of such terms.

We now demonstrate that Eq.~(\ref{eq:solrho}) is actually the
transverse displacement, giving the equilibrium transverse
positions of the zigzag structure, by verifying that
Eq.~(\ref{eq:solrho}), together with $\bar\Psi^{(\pm)}_{\delta
k}=0$ for $\delta k>0$, yields a stable solution.  To check
stability the matrix of the second derivatives of $V_{\rm eff}$
with respect to the various variables should be calculated. The
second derivative of $V_{\rm eff}$, given in Eq.~(\ref{V:eff}).
with respect to $\bar\varrho$ is positive,
\begin{eqnarray} 
\left . \frac{\partial^2V_{\rm
eff}}{\partial\varrho^{2}}\right|_{\{\varrho,\Psi_{\delta
k}\}=\{\bar\varrho,0\}}&=&-2 m\beta_0>0\,. \end{eqnarray} In order to
investigate the coupling of the soft mode with the modes with
$\delta k\neq 0$, one can calculate the second derivatives of
$V_{\rm eff}$ with respect to $\Psi_0^\sigma$. We find
\begin{eqnarray} \left.\frac{\partial^2V_{\rm eff}}{\partial
\Psi^\sigma_0\partial \Psi_{\delta
k}^{\sigma'(\pm)}}\right|_{\{\varrho,\Psi_{\delta
k}\}=\{\bar\varrho,0\}}&=&0\:. \nonumber \end{eqnarray} This result
shows that the derivatives with respect to $\Psi_0^\sigma$ form a
sub-block of the stability matrix that can be diagonalized
separately. All its eigenvalues are found to be positive. The
other second derivatives at these points read 
\begin{eqnarray}
\left . \frac{\partial^2V_{\rm eff}}{\partial
\Psi^{z(\pm)2}_{\delta k}}\right|_{\{\varrho,\Psi_{\delta
k}\}=\{\bar\varrho,0\}}&=&m\beta_{\delta k}
+8A\bar{\varrho}^2+16A\bar\Psi_0^{z2}\;,
\\
\left . \frac{\partial^2V_{\rm eff}}{\partial
\Psi^{y(\pm)2}_{\delta k}}\right|_{\{\varrho,\Psi_{\delta
k}\}=\{\bar\varrho,0\}}&=&m\beta_{\delta k}
+8A\bar{\varrho}^2+16A\bar\Psi_0^{y2}\;,
\\
\left . \frac{\partial^2V_{\rm eff}}{\partial
\Psi^{y(\pm)}_{\delta k}\partial \Psi^{z(\mp)}_{\delta
k}}\right|_{\{\varrho,\Psi_{\delta k}\}=\{\bar\varrho,0\}}&=&0\;,
\\
\left . \frac{\partial^2V_{\rm eff}} {\partial
\Psi^{y(\pm)}_{\delta k} \partial \Psi_{\delta
k}^{z(\pm)}}\right|_{\{\varrho,\Psi_{\delta
k}\}=\{\bar\varrho,0\}}&=& 16A\bar{\Psi}_0^y\bar{\Psi}_0^z\;,
 \end{eqnarray} where we have used that $0>\beta_{\delta
k}>\beta_0$. This result shows that the modes
$\Psi^{y(\pm)}_{\delta k}$ and $\Psi^{z(\pm)}_{\delta k}$ are
coupled in pairs. All contributions resulting of differentiation
of the function ${\mathcal F}$ in Eq.~(\ref{V:4:soft}) vanish. The
stability matrix splits into 2$\times$2 blocks that can be
diagonalized separately. Using Eq.~(\ref{eq:solrho}) and
$\beta_0<\beta_{\delta k}$ one finds that the eigenvalues of each
block are $m(\beta_{\delta k}-2\beta_0)$ and $m(\beta_{\delta
k}-6\beta_0)$, hence both positive. Therefore, a gap opens between
the soft mode frequency and the frequency of the modes at $\delta
k\neq 0$ in the vicinity of the transition point. Therefore, the
instability is driven by the soft modes with wave vector $k_0$,
determining the order of the zigzag
phase~\cite{Mukamel86,Mukamel87}.

\subsection{Behaviour at the critical point}

From Eq.~(\ref{V:eff}), using the results of the previous section
we can now write the effective potential for the soft modes, which
reads \begin{eqnarray} \label{V:soft} V^{\rm soft} ={\mathcal
V}\left(\left(\Psi_0^{y}\right)^2+\left(\Psi_0^{z}\right)^2\right)+A\left(\left(\Psi_0^{y}\right)^2+\left(\Psi_0^{z}\right)^2\right)^2,
\end{eqnarray} where $A$ is given by Eq.~\eqref{eq:A} and
\begin{equation} \mathcal
V=\frac{m}{2}\beta_0=\frac{1}{2}m\left(\nu_t^2-\nu_t^{(c)2}\right).
\end{equation} Here, we have used that $\beta_0=\omega_{\perp,{\rm
min}}^2$, which in turn is given by Eq.~(\ref{w:perp:min}). Hence,
for $\mathcal V>0$ the potential $V^{\rm soft}$ has a single
minimum with $\Psi_0^{\sigma}=0$, and the linear chain is the
ground state structure, while for $\mathcal V<0$ the potential
landscape has the characteristic form of a Mexican hat with
degenerate zigzag ground states at different angles around the
symmetry axis. Indeed, while the order parameter $\bar\varrho$ is
fixed by condition \eqref{eq:solrho}, the ratio
$\bar\Psi^y_0/\bar\Psi^z_0$ is arbitrary. The system hence
possesses "Goldstone modes" at zero frequency, which are a
consequence of the symmetry by rotations around the trap axis.

The transverse displacement from the trap is given from
Eq.~(\ref{eq:solrho}) by using Eq.~(\ref{soft:mode}) and the
relation $b/2=\bar\varrho/\sqrt{N}$, which links the displacement
in real space with its Fourier decomposition. Hence, for
$\nu_t<\nu_t^{(c)}$ the transverse displacement from the trap
depends on $\nu_t$ as \begin{equation} \label{eq:b}
b=\bar
b\sqrt{\nu_t^{(c)}-\nu_t}\;, \end{equation} with $\bar
b=\sqrt{2m\nu_t^{(c)}/\mathcal A}$. This behaviour is in agreement with the
numerical results in~\cite{Schiffer93}.

From Eq.~\eqref{eq:b} we evaluate the difference between the ground state energy of the linear and of the zigzag structure. Considering the energy per particle, from Eq.~\eqref{V:soft} we find 
\begin{eqnarray}
&&\Delta E = \frac{V^{\rm soft}(\nu_t\to\nu_t^{(c)-})-V^{\rm soft}(\nu_t\to\nu_t^{(c)+})}{N}
\nonumber\\
&&=-\frac 12 m \mathcal C a^2 (\nu_t-\nu_t^{(c)})^2
\end{eqnarray}
where $\mathcal C = 112\;\zeta(3)/[93\;\zeta(5)]$,
and whose second derivative with respect to $\nu_t$ is clearly discontinuous at the critical point.
This result is consistent with the result presented in \cite{Piacente2005}, where a discontinuity in the second derivative of the ground state energy with respect to the particles density was found. 

\subsection{Discussion}

Using symmetry arguments we have demonstrated that the transition
from a linear chain to a zigzag structure, in a system of
anisotropically confined charges, is a second-order phase
transition, whose order parameter is the displacement from the
trap axis. This theory has been developed in the thermodynamic
limit, fixing the interparticle distance $a$ as the number of ions
was let to infinity. In this limit, we found that the soft modes
are the zigzag modes of the linear chain, whose periodicity is
equal to twice the interparticle distance $a$. The instability is
thus driven by these modes as the transverse potential is changed
across the critical value $\nu_t^{(c)}$.

These considerations are strictly valid for $N\to \infty$, but can
still be useful for finite systems, and in particular when the
ions are confined in a trap, which provides also axial harmonic
confinement. While detailed quantitative predictions can be only made by
accurately evaluating the finite-size corrections, we can still
make some reasonable conjectures, based on previous results in the
literature and on our theory. Inside a harmonic trapping
potential, the interparticle distance between the ions varies
along the chain and it is minimal at the center. Numerical
results, based on molecular dynamics simulations, showed that in
this case the zigzag structure appears at the center of the
chain where the density is highest~\cite{Schiffer93}. Analytical studies found that the short
wavelength modes are characterized by largest displacements at the
chain center, while the ions at the edge almost do not
move~\cite{morigi-pre}. In this case, hence, we can still identify
the zigzag mode of the ion chain with the soft mode. In the
presence of axial confinement, however, both the transverse as
well as the axial equilibrium points will change. In particular,
when going to the zigzag structure the axial density of ions in
the center will increase. Close to the transition point, one finds
that the axial corrections to the linear chain positions are much
smaller than the transverse displacements from the trap axis as this is a quantity that follows the order parameter. This
is also confirmed by the analysis made for the simple case of
three ions in Sec.~\ref{Sec:3:ions}, where close to the critical
value of the aspect ratio the transverse displacement varies
faster than the axial one. Therefore, we expect that our theory
will still provide reasonable predictions close to the critical
point, also in presence of axial confinement.

\section{Conclusions and outlook} \label{Sec:Conclusions}

The structural phase transition from a linear chain to a zigzag
configuration, in a system composed of trapped singly-charged
particles, is a second order phase transition. Using a mean field
approach we have derived a classical model, describing the system
at the critical point and its vicinity. Our theory is analytical
and its predictions agree with the numerical simulations
of~\cite{Schiffer93} and~\cite{Piacente2005}.

\begin{figure} \includegraphics{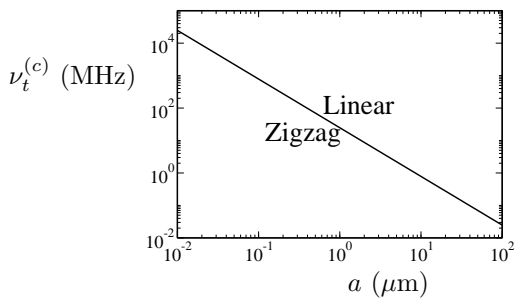}
\caption{\label{Phase}Phase diagram close to the linear-zigzag
transition in the thermodynamic limit, for $^{40}$Ca$^{+}$ ions. The horizontal axis is
the interparticle spacing $a$ in $\mu$m, the vertical axis the
corresponding critical frequency $\nu_t^{(c)}$ in MHz. This
graphic does not report further curves in the left region, giving
the transition to more complex structures. A detailed study of the
transitions to these structures can be found in~\cite{MPQ1,MPQ2}
and~\cite{Piacente2005}.} \end{figure}

The corresponding phase diagram is shown in Fig.~\ref{Phase}, it
shows the regions of stability of the linear chain as a function of
the interparticle spacing $a$ and the transverse frequency
$\nu_t$. The phase diagram is evaluated in the thermodynamic
limit, corresponding in keeping $a$ fixed as $N$ and the chain
length go to infinity. The analysis is valid for $T=0$, where long-range order in one-dimensional structures
exists. 
The quantum statistics
of the particles at these densities seem irrelevant even at these ultralow temperatures
since the
interaction energy at all stages much larger than the kinetic
energy, and the particles can be considered distinguishable at all
effects~\cite{Javanainen}. On the other hand, at the critical
point, where large fluctuations of the transverse motion
classically occur, quantum effects may be relevant and could be in
principle observed. 

%%%%%%%%%%%%%%%%%%%%%%%%%
%%%%%%%%%%%%%%%%%%%%%%%%%

\acknowledgments The authors thank Efrat Shimshoni, Grigory
Astrakharchik, Eugene Demler, Bert Halperin, and Tommaso Roscilde
for stimulating discussions and useful comments. This work was
partly supported by the European Commission (EMALI,
MRTN-CT-2006-035369; SCALA, Contract No.\ 015714; QOQIP,
MOIF-CT-2005-8688), by the Spanish Ministerio de Educaci\'on y
Ciencia (Consolider Ingenio 2010 "QOIT", CSD2006-00019; QLIQS,
FIS2005-08257; Ramon-y-Cajal individual fellowship), and by the
National Science Foundation through a grant to the Institute for
Theoretical Atomic, Molecular, and Optical Physics at the
Smithsonian Center for Astrophysics and Harvard Department of
Physics. S.F. acknowledges the US-Israel Binational Science Foundation (BSF), The Israel Science Foundation (ISF), the Minerva Center of Nonlinear Physics of Complex Systems, and the fund for Promotion of Research at the Technion. 
G.M. thanks Herbert Walther, who motivated this work.

\begin{appendix}

\section{Expansion about the equilibrium posisions of the linear chain}
\label{App:A}

In this Appendix we evaluate the higher order terms of the
expansion of the potential in Eq.~(\ref{Eq:potential}) about the
equilibrium position of the linear chain. For this purpose we
rewrite the interparticle distance as $|{\bf r_{i}}-{\bf
r_j}|=\sqrt{A_{ij}+\tau_{ij}+\epsilon_{ij}}$ with
\begin{eqnarray*}
&&A_{ij}=(i-j)^2a^2,\\
&&\tau_{ij}=2a(i-j)(q_i-q_j)\;,\\
&&\epsilon_{ij}=(q_i-q_j)^2+\delta_{ij}^2\;, \end{eqnarray*} and
$\delta_{ij}^2=(y_i-y_j)^2+(z_i-z_j)^2$ and we have used
$x_j=ja+q_j$. We now expand in the parameters $\epsilon_{ij}$ and
$\tau_{ij}$, assuming that they are small with respect to
$A_{ij}$, i.e., to the axial equilibrium distances between the
ions when the chain is stable. We will check later for consistency
of this assumption. We can write \begin{equation}
V^{(l)}=\frac{Q^2}{2}\sum_{i,j\neq i} W_{ij}^{(l)}\;,
\end{equation} with \begin{eqnarray*}
&&W_{ij}^{(0)}=\frac{1}{|i-j|a}\;,\\
&&W_{ij}^{(1)}=-\frac{\sigma_{ij}}{(i-j)^2a^2}(q_i-q_j),\\
&&W_{ij}^{(2)}=\frac{1}{2|i-j|^3a^3}(2(q_i-q_j)^2-\delta_{ij}^2),\\
&&W_{ij}^{(3)}=\frac{\sigma_{ij}}{2|i-j|^4a^4}(q_i-q_j)\left[3\delta_{ij}^2-2(q_i-q_j)^2\right],\\
&&W_{ij}^{(4)}=\frac{1}{|i-j|^5a^5}\Bigl(\frac{3}{8}\delta_{ij}^4+(q_i-q_j)^4-3\delta_{ij}^2(q_i-q_j)^2\Bigr),\end{eqnarray*}
where we have introduced $\sigma_{ij}=(i-j)/|i-j|$. We notice that
\begin{eqnarray} V^{(1)}=-\frac{Q^2}{2a^2}\sum_i\sum_{j\neq
i}\sigma_{ij}\frac{q_i-q_j}{(i-j)^2}=0 \end{eqnarray} as one can
easily verify by using the definition of $\sigma_{ij}$. This is
satisfied also in the ion chain in presence of an axial trapping
potential, since $V^{(1)}=0$ determines the equilibrium positions.

\section{Potential terms for the normal modes of the linear chain}
\label{app:3d}

Here we report the third and fourth order terms of the potential,
decomposed into the modes
$\Theta_{k}=\Theta_k^{(+)}-i\Theta_k^{(-)}$,
$\Psi_k^\sigma=\Psi_k^{\sigma(+)}-i\Psi_k^{\sigma(-)}$.  The third
order term takes the form \begin{eqnarray} V^{(3)}&=&
\sum_{k_1+k_2+k_3=0} B(k_1,k_2,k_3)
\label{eq:V:3}\\
&
&\times\Bigl[\Theta_{k_1}\left(3\sum_{\sigma=y,z}\Psi_{k_2}^{\sigma}\Psi_{k_3}^{\sigma}-2\Theta_{k_2}\Theta_{k_3}\right)\,\nonumber
\end{eqnarray} where the sum runs over positive and negative
values of $k_j$ and \begin{eqnarray}
B(k_1,k_2,k_3)&=&-i\sqrt{\frac{2}{N}}\frac{Q^2}{a^4}\sum_{m>0}
\frac{1}{m^4} \prod_{p=1}^3\sin{\frac{k_p m a}{2}}\;.
\end{eqnarray} Term~(\ref{eq:V:3}) is real, as it is visible by
using the decomposition into even and odd modes. In particular, it
has odd parity, coupling either three odd modes or two odd modes
with an even one.

The quartic term reads \begin{eqnarray} \label{eq:V:4}
V^{(4)}&=&\sum_{k_1+k_2+k_3+k_4=0}A(k_1,k_2,k_3,k_4)\\
&\times&\left[
\frac{3}{8}\sum_{\sigma,\sigma'=y,z}\Psi^\sigma_{k_1}\Psi^\sigma_{k_2}\Psi^{\sigma'}_{k_3}\Psi^{\sigma'}_{k_4}
+\Theta_{k_1}\Theta_{k_2}\Theta_{k_3}\Theta_{k_4} \right .\nonumber\\
&-&\left .
3\sum_{\sigma=y,z}\Psi^\sigma_{k_1}\Psi^\sigma_{k_2}\Theta_{k_3}\Theta_{k_4}\right]
\nonumber \end{eqnarray} with \begin{eqnarray} \label {eq:Ak1}
&&A(k_1,k_2,k_3,k_4)=\frac{4}{N}\frac{Q^2}{a^5}\sum_{m>0}\frac{1}{m^5}
\prod_{p=1}^4\sin{\frac{k_p m a}{2}} \end{eqnarray} This term is
even, and it thus couples either four modes with the same parity,
or two odd modes with two even ones.

\end{appendix}
\bibliography{ionchain}
\end{document}